# Service-Aware Hosting Capacity Assessment of Edge Data Centers with Workload Flexibility

Linhan Fang[1], Haoxiang Wan[1], Anurag Srivastava[2], and Xingpeng Li[1]

1. Department of Electrical and Computer Engineering, University of Houston, Houston, TX, USA
2. Lane Department of Computer Science and Electrical Engineering, West Virginia University, Morgantown, WV, USA

***Abstract*—The growing demand for edge computing, content delivery, and AI services is driving the interconnection of edge data centers (DCs) with distribution networks. Before interconnection, utilities evaluate how much DC capacity can be accommodated without violating feeder current and voltage limits. Fixed peak load representations can overlook service dependent demand patterns, cooling related facility demand, and delay tolerant computing tasks. This paper develops a hosting capacity (HC) assessment model for edge DCs with shiftable workload scheduling. Three representative types are considered: mobile edge computing (MEC), content delivery network (CDN), and AI DCs. Facility demand is modeled by combining fixed base demand, service driven demand, prescribed cooling-overhead coefficient, and a shiftable workload component. Workload execution is optimized under arrival causality, maximum delay, daily energy balance constraints, and the resulting active and reactive loads are embedded in a distribution power flow model. Compared with the baseline without load shifting, the maximum bus level HC gain on the IEEE 33-bus system reaches 4.46% for MEC DCs and 9.89% for AI DCs, while CDN case shows negligible gain under the assumed cache management windows.**

## I. Introduction

The rapid growth of data centers (DCs), driven by cloud services, edge computing, and AI workloads, is increasing stress on the electrical grid planning and interconnection processes. Recent studies report a growing electricity demand from data centers in the United States, with further increases expected from high density computing and advanced cooling requirements [1]-[2]. In addition, the concentrated siting of AI and cloud facilities may create regional grid stress and require earlier interconnection studies and infrastructure upgrades [3]. Data center deployment has therefore become a grid planning problem, especially when new facilities are connected to constrained distribution networks.

Before an edge data center is connected, its hosting capacity (HC) should be evaluated to determine the maximum facility size that can be accommodated without violating network operating limits. In distribution systems, HC is mainly constrained by voltage regulation, branch thermal limits, and power flow feasibility [4]-[5]. For data center integration, this value also depends on feeder topology, electrical distance from the substation, local reactive support, and possible network reinforcement [6]. Facility level factors, including cooling efficiency and reliability requirements, further affect the active and reactive demand seen by the grid [7]. These factors make HC assessment a necessary step for both interconnection screening and distribution planning.

Edge data centers are not uniform loads. Their demand patterns depend strongly on the service function. Mobile edge computing facilities support latency sensitive applications such as task offloading, mobile traffic processing, and local analytics [8]-[9]. Content delivery network-oriented edge facilities serve content requests and rely on caching, request routing, content replication, and background update mechanisms that may be scheduled over selected time windows [10]-[11]. Edge AI and cloudlet data centers combine real time inference with batch-oriented computing tasks such as model updates, batch inference, and index rebuilding [12]-[13]. These differences lead to distinct load profiles and varying levels of temporal flexibility, which are not captured by a single fixed load curve.

Existing studies have shown that flexible resources can improve data center HC and reduce grid impacts. Battery storage, dispatchable generation, and reactive power support can relieve voltage or thermal constraints in active distribution networks [14]. Coordinated planning of reinforcement and storage can further improve the efficiency of distribution network expansion [6]. At the grid system level, spatial flexibility of data center workloads can reduce congestion, renewable curtailment, and operating cost [15], while energy storage can support system resilience under stressed operating conditions [16]-[19]. However, these approaches mainly rely on grid side resources or spatial load transfer.

This paper studies internal shiftable workload scheduling as a complementary source of flexibility for edge data center interconnection. Many computing tasks, especially batch workloads, do not need to be executed immediately after arrival [20]. Their execution can be shifted within an allowable delay window while preserving the total workload energy. When shifted execution reduces demand during feeder constrained hours, the same facility type may achieve a higher feasible interconnection capacity. To quantify this effect, this paper develops a HC assessment model that represents three edge DC archetypes, incorporates a prescribed cooling overhead coefficient, and schedules shiftable workloads subject to causality, delay, and daily energy balance constraints. The resulting active and reactive data center demands are embedded in a distribution network power flow model with voltage and branch current limits.

## II. Methodology and Models

This section presents the HC assessment model for an edge data center connected at a candidate bus in a distribution network. For each simulation case, the candidate bus and the edge data center type are fixed, and the model maximizes the facility level HC. The methodology consists of three steps. First, workload arrival profiles are constructed for representative edge data center service types based on their service archetypes. Second, the DC facility demand is constructed by combining fixed base load, service-driven facility load, prescribed cooling-overhead profile, and shiftable workload. Third, the resulting active and reactive data center demands are embedded in the distribution network power flow constraints.

### A. Edge DC Workload Archetypes and Shiftable Workload Arrival Profiles

Three representative edge data center types are considered: mobile edge computing (MEC), content delivery network (CDN), and AI data centers. They are not intended to represent universal measured traces. Instead, they provide reproducible workload archetypes for comparing how different service functions affect HC. The key modeling difference among them lies in the temporal arrival pattern of shiftable workloads with tolerance for delays.

Let $\mathcal{K} = \{\text{MEC}, \text{CDN}, \text{AI}\}$ denote the set of edge data center archetypes, and let $\mathcal{T}$ denote the hourly scheduling horizon. For each type $k \in \mathcal{K}$, $\rho_{k,t}^{\text{DC}}$ denotes the peak normalized facility load profile, and $\omega_{k,t}^{\mathcal{B}}$ represents the relative arrival weight of shiftable workload. These profiles satisfy:

$$\rho_{k,t}^{\text{DC}} \geq 0, \max_{t \in \mathcal{T}} \rho_{k,t}^{\text{DC}} = 1, \forall t \in \mathcal{T}, \forall k \in \mathcal{K} \quad (1)$$

$$\omega_{k,t}^{\mathcal{B}} \geq 0, \max_{t \in \mathcal{T}} \omega_{k,t}^{\mathcal{B}} = 1, \forall t \in \mathcal{T}, \forall k \in \mathcal{K} \quad (2)$$

For the MEC DC archetype, the shiftable workload arrival profile is assumed to be correlated with mobile service activity, reflecting background computing tasks associated with edge services driven traffic. For the CDN DC archetype, real-time content serving is treated as non-shiftable, while cache refresh, prefetching, synchronization, backup, and maintenance tasks are assigned to predefined cache management windows. For the AI DC archetype, the shiftable workload arrival profile represents a hybrid pattern that combines scheduled batch operations with tasks correlated to services.

### B. Facility Load and Temporal Shifting Model

This section converts the representative profiles defined in Section II-A into the active and reactive demand of an edge DC. The model describes hourly demand at the facility level and is used for HC assessment. It does not attempt to represent individual servers, job queues, or internal cooling control. The facility demand is divided into a fixed facility component and an IT dependent component. The fixed component represents demand that is not affected by workload scheduling, such as auxiliary equipment, networking devices, control systems, and minimum idle operation. It is modeled as a constant fraction of the facility level HC at bus $b$ of the set of candidate buses $B$.

$$P_{b,k}^{\text{DC,base}} = \lambda_{base} H_{b,k}^{\text{DC}}, \forall b \in B, \forall k \in \mathcal{K} \quad (3)$$

The hourly cooling coefficient $\eta_t^{\text{cooling}}$ is treated as a prescribed input associated with the adopted outdoor temperature profile. It represents the additional facility demand associated with IT power consumption. The baseline IT demand before workload shifting is defined in (4), and the corresponding baseline facility demand is given in (5). Since $\rho_{k,t}^{\text{DC}}$ represents the normalized facility load profile, (4) infers the corresponding IT demand under the prescribed cooling-overhead coefficient, while (13) evaluates how shifted IT execution changes the total facility load. This construction preserves the interpretation of $H_{b,k}^{\text{DC}}$ as the peak facility load.

$$P_{b,k,t}^{\text{IT,pre}} = (1 - \lambda_{base}) H_{b,k}^{\text{DC}} \cdot \frac{\rho_{k,t}^{\text{DC}}}{\left(1 + \eta_t^{\text{cooling}}\right)}, \forall t \in \mathcal{T}, \forall k \in \mathcal{K}, \forall b \in B \quad (4)$$

$$P_{b,k,t}^{\text{DC,pre}} = P_{b,k}^{\text{DC,base}} + \left(1 + \eta_t^{\text{cooling}}\right) P_{b,k,t}^{\text{IT,pre}}, \forall t \in \mathcal{T}, \forall k \in \mathcal{K}, \forall b \in B \quad (5)$$

The shiftable workload $A_{b,k,t}^{\mathcal{B}}$ is modeled as a component of the baseline IT demand. Equation (6) assigns a fraction $\beta_k$ of the daily IT energy to hourly shiftable workload arrivals according to the relative arrival weight $\omega_{k,t}^{\mathcal{B}}$. Since $\omega_{k,t}^{\mathcal{B}}$ is used as a peak normalized weight profile, it is divided by its daily sum when shiftable workload arrivals are constructed.

$$A_{b,k,t}^{\mathcal{B}} \Delta t = \beta_k \left( \sum_{\tau \in \mathcal{T}} P_{b,k,\tau}^{\text{IT,pre}} \Delta t \right) \cdot \left( \omega_{k,t}^{\mathcal{B}} / \sum_{\tau \in \mathcal{T}} \omega_{k,\tau}^{\mathcal{B}} \right), \forall t \in \mathcal{T}, \forall k \in \mathcal{K}, \forall b \in B \quad (6)$$

The remaining IT demand that cannot be shifted is defined in (7). For all reported cases, the selected values of $\beta_k$ and the shiftable workload arrival weights satisfy $A_{b,k,t}^{\mathcal{B}} \leq P_{b,k,t}^{\text{IT,pre}}$, so the residual IT demand remains nonnegative.

$$P_{b,k,t}^{\text{IT,NF}} = P_{b,k,t}^{\text{IT,pre}} - A_{b,k,t}^{\mathcal{B}}, \forall t \in \mathcal{T}, \forall k \in \mathcal{K}, \forall b \in B \quad (7)$$

Let $S_{b,k,t}^{\mathcal{B}}$ denote the shiftable workload executed at hour $t$. The actual IT demand after scheduling is obtained by adding the residual IT demand and the scheduled workload execution.

$$S_{b,k,t}^{\mathcal{B}} \geq 0, \forall t \in \mathcal{T}, \forall k \in \mathcal{K}, \forall b \in B \quad (8)$$

$$P_{b,k,t}^{\text{IT,sch}} = P_{b,k,t}^{\text{IT,NF}} + S_{b,k,t}^{\mathcal{B}}, \forall t \in \mathcal{T}, \forall k \in \mathcal{K}, \forall b \in B \quad (9)$$

The temporal feasibility of shiftable workload scheduling is enforced by (10)-(12). Constraint (10) prevents execution before arrival. Constraint (11) requires shiftable workload to be completed within the allowed delay window. Constraint (12) preserves the total daily shiftable workload energy.

$$\sum_{\ell=1}^{t} S_{b,k,\ell}^{\mathcal{B}} \Delta t \leq \sum_{\ell=1}^{t} A_{b,k,\ell}^{\mathcal{B}} \Delta t, \forall t \in \mathcal{T}, \forall k \in \mathcal{K}, \forall b \in B \quad (10)$$

$$\sum_{\ell=1}^{t} S_{b,k,\ell}^{\mathcal{B}} \Delta t \geq \sum_{\ell=1}^{t-D_k^{\max}} A_{b,k,\ell}^{\mathcal{B}} \Delta t, \forall t > D_k^{\max}, \forall k \in \mathcal{K}, \forall b \in B \quad (11)$$

$$\sum_{t \in \mathcal{T}} S_{b,k,t}^{\mathcal{B}} \Delta t = \sum_{t \in \mathcal{T}} A_{b,k,t}^{\mathcal{B}} \Delta t, \forall k \in \mathcal{K}, \forall b \in B \quad (12)$$

where $D_k^{\max}$ is the maximum allowable delay duration for data center type $k$. A cyclic daily horizon is adopted to link the end and beginning of the scheduling window.

The active data center demand passed to the grid is constructed in (13) by combining the fixed facility base load with the cooling-adjusted actual IT load. The corresponding reactive demand is calculated in (14) using a fixed power factor angle $\theta$. Besides, (15) is enforced to strictly bound the active data center demand within its predefined peak capacity limit.

$$P_{b,k,t}^{\text{DC}} = P_{b,k}^{\text{DC,base}} + \left(1 + \eta_t^{\text{cooling}}\right) P_{b,k,t}^{\text{IT,sch}}, \forall t \in \mathcal{T}, \forall k \in \mathcal{K}, \forall b \in B \quad (13)$$

$$Q_{b,k,t}^{\text{DC}} = P_{b,k,t}^{\text{DC}} \tan\theta, \forall t \in \mathcal{T}, \forall k \in \mathcal{K}, \forall b \in B \quad (14)$$

$$0 \leq P_{b,k,t}^{\text{DC}} \leq H_{b,k}^{\text{DC}}, \forall t \in \mathcal{T}, \forall k \in \mathcal{K}, \forall b \in B \quad (15)$$

### C. System Constraints

The active and reactive data center demands defined in (13) and (14) are assigned to the selected candidate bus. For each simulation case, the optimization problem is solved independently for each fixed candidate bus $b$ and DC archetype $k$.

$$\delta_{i,b} = \begin{cases} 1, & i = b \\ 0, & i \neq b \end{cases} \tag{16}$$

For each fixed candidate bus and DC archetype, the HC is obtained by maximizing the admissible facility capacity:

$$\text{objective} = \max H_{b,k}^{\text{DC}} \tag{17}$$

where $H_{b,k}^{\text{DC}}$ is the maximum edge DC capacity that can be connected at bus $b$ for DC type $k$ without violating network operating limits.

The active and reactive power balance equations for non-substation buses $\mathcal{N}$ and substation bus $\mathcal{G}$ are given in (18)-(19).

$$\begin{cases} \sum_{m\in\mathcal{U}(i)} (P_{mi}^t - r_{mi} \cdot l_{mi}^t) = \sum_{j\in\mathcal{D}(i)} P_{ij}^t + P_{\text{load},i}^t + \delta_{i,b} P_{b,k,t}^{\text{DC}} \\ \sum_{m\in\mathcal{U}(i)} (Q_{mi}^t - x_{mi} \cdot l_{mi}^t) = \sum_{j\in\mathcal{D}(i)} Q_{ij}^t + Q_{\text{load},i}^t + \delta_{i,b} Q_{b,k,t}^{\text{DC}} \end{cases} \tag{18}$$
$$,\forall i \in \mathcal{N}, \forall t \in \mathcal{T}$$

$$\begin{cases} \sum_{m\in\mathcal{U}(g)} (P_{mg}^t - r_{mg} \cdot l_{mg}^t) = -P_g^t + \sum_{j\in\mathcal{D}(g)} P_{gj}^t + P_{\text{load},g}^t \\ \sum_{m\in\mathcal{U}(g)} (Q_{mg}^t - x_{mg} \cdot l_{mg}^t) = -Q_g^t + \sum_{j\in\mathcal{D}(g)} Q_{gj}^t + Q_{\text{load},g}^t \end{cases} \tag{19}$$
$$,\forall g \in \mathcal{G}, \forall t \in \mathcal{T}$$

where for each node $i \in \mathcal{N}$, $\mathcal{D}(i)$ and $\mathcal{U}(i)$ represent the sets of its downstream and upstream nodes, respectively. The DC load terms are added only at the selected installation bus through the indicator in (16).

The branch voltage relationship is modeled using the distribution flow voltage drop equation in (20). Equation (21) provides the SOCP relaxation for the relation among branch current, power flow, and bus voltage in the radial feeder model. The substation voltage is fixed by (22).

$$v_i^t - v_j^t = 2\left(r_{ij}P_{ij}^t + x_{ij}Q_{ij}^t\right) - \left(r_{ij}^2 + x_{ij}^2\right)l_{ij}^t$$
$$,\forall (i,j) \in \mathcal{L}, \forall t \in \mathcal{T} \tag{20}$$

$$v_i^t \cdot l_{ij}^t \geq (P_{ij}^t)^2 + (Q_{ij}^t)^2, \forall (i,j) \in \mathcal{L}, \forall t \in \mathcal{T} \tag{21}$$

$$v_g^t = (V_g^{\text{ref}})^2, \forall g \in \mathcal{G}, \forall t \in \mathcal{T} \tag{22}$$

The nodal voltage and branch current limits are enforced as:

$$v^{\min} \leq v_i^t \leq v^{\max}, \forall i \in \mathcal{N}, \forall t \in \mathcal{T} \tag{23}$$

$$0 \leq l_{ij}^t \leq (I_{ij}^{\text{lim}})^2, \forall (i,j) \in \mathcal{L}, \forall t \in \mathcal{T} \tag{24}$$

where $v^{\min}$ and $v^{\max}$ are the square of minimum and maximum per-unit voltage, and $I_{ij}^{\text{lim}}$ represents the current rating between node $i$ and $j$.

## III. Case Studies

This section evaluates the proposed HC model on the IEEE 33-bus radial distribution system for MEC, CDN, and AI DCs. The 32 non-substation buses are considered as candidate locations. Each case uses a 24-hour horizon with $\Delta t$ =1 h, voltage limits of 0.95-1.05 p.u., standard branch current limits, a DC power factor of 0.98, and a fixed base load ratio of $\lambda_{base}$=0.2. A fixed daily outdoor temperature profile and the corresponding hourly cooling coefficient $\eta_t^{\text{cooling}}$ are used for the cooling-aware load calculation. All models are solved in Gurobi as continuous conic programs with a 10 MVA system base power and a relative optimality tolerance of 0.1%.

### *A. Case Setup and Workload Profiles*

Fig. 1 shows the peak normalized facility load profiles and the relative shiftable workload arrival profiles used in the case studies. These profiles are representative scenario inputs used to compare how the timing of service demand and shiftable workload affects HC. The MEC DC follows a diurnal facility load profile driven by traffic, and its shiftable workload arrival weight is correlated to mobile service activity. The CDN DC represents content serving demand with scheduled cache management windows for tasks such as refresh, prefetching, synchronization, backup, and maintenance. The load profile of an AI data center features high utilization, driven by a shiftable workload arrival pattern that combines scheduled batch operations with tasks correlated to specific services.

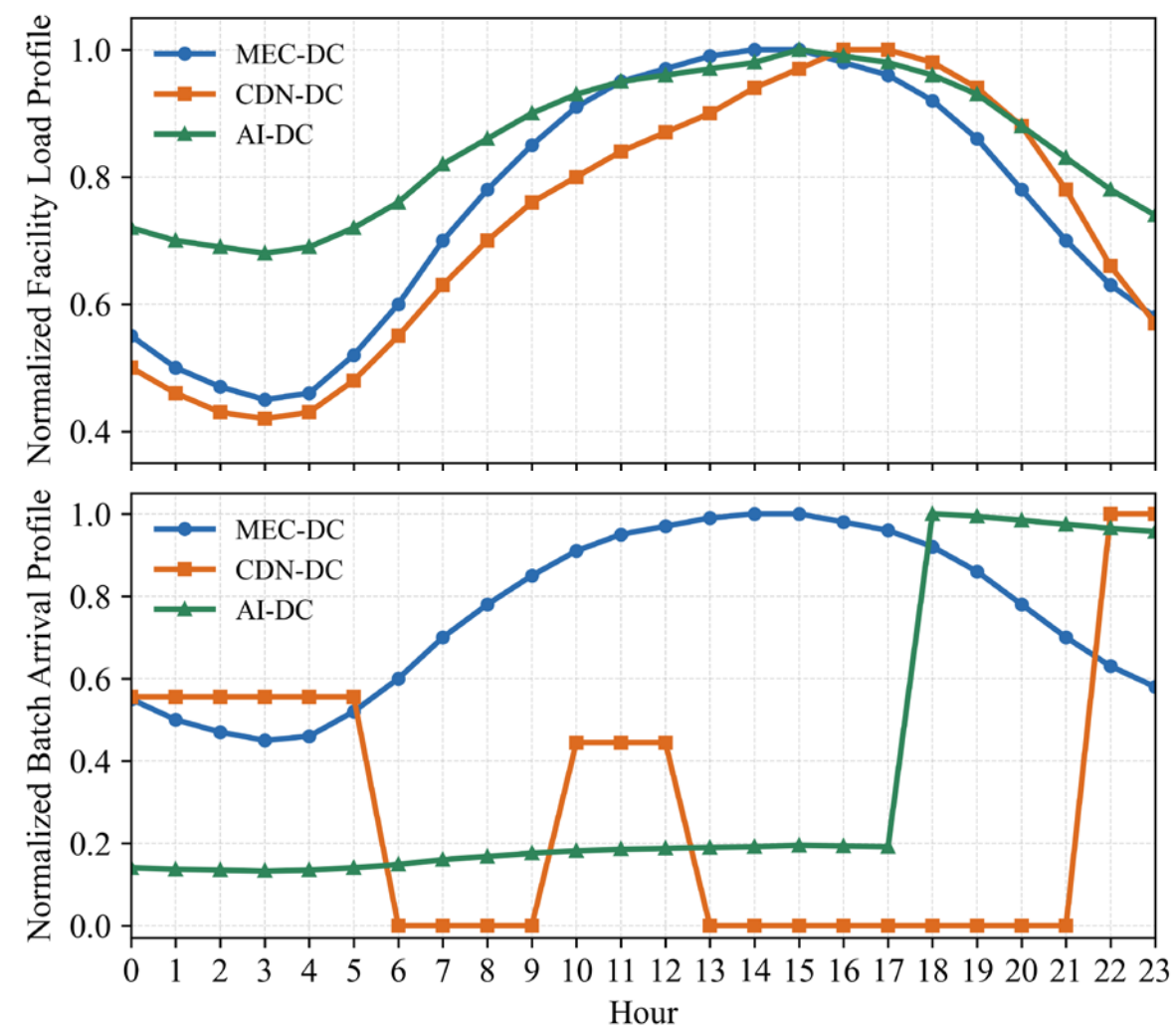

Fig. 1. Representative edge DC profiles: (a) Peak normalized facility load profiles. (b) Peak normalized shiftable workload arrival weight profiles.

Table I summarizes the main and sensitivity settings for the shiftable workload ratio $\beta_k$ and the maximum delay tolerance $D_k^{\max}$. MEC DC is assigned a smaller shiftable ratio and shorter delay window because of their latency sensitive service requirements. CDN DC has moderate flexibility associated with cache refresh, prefetching, synchronization, and maintenance. AI DC is assigned a larger flexibility range because AI workloads may contain a higher share of batch-oriented computing tasks. These settings are used in the following HC comparisons.

### *B. MEC DC Results: Traffic Correlated Delay Flexibility*

The MEC DC case shows a moderate HC improvement under the tested flexibility settings. As shown in Fig. 2, increasing the shiftable workload ratio $\beta_k$ improves HC, and a longer delay window $D_k^{\max}$ provides additional benefit. The effect of $D_k^{\max}$ becomes more visible when $\beta_k$ increases, because more background workload can be redistributed away from feeder stress hours.

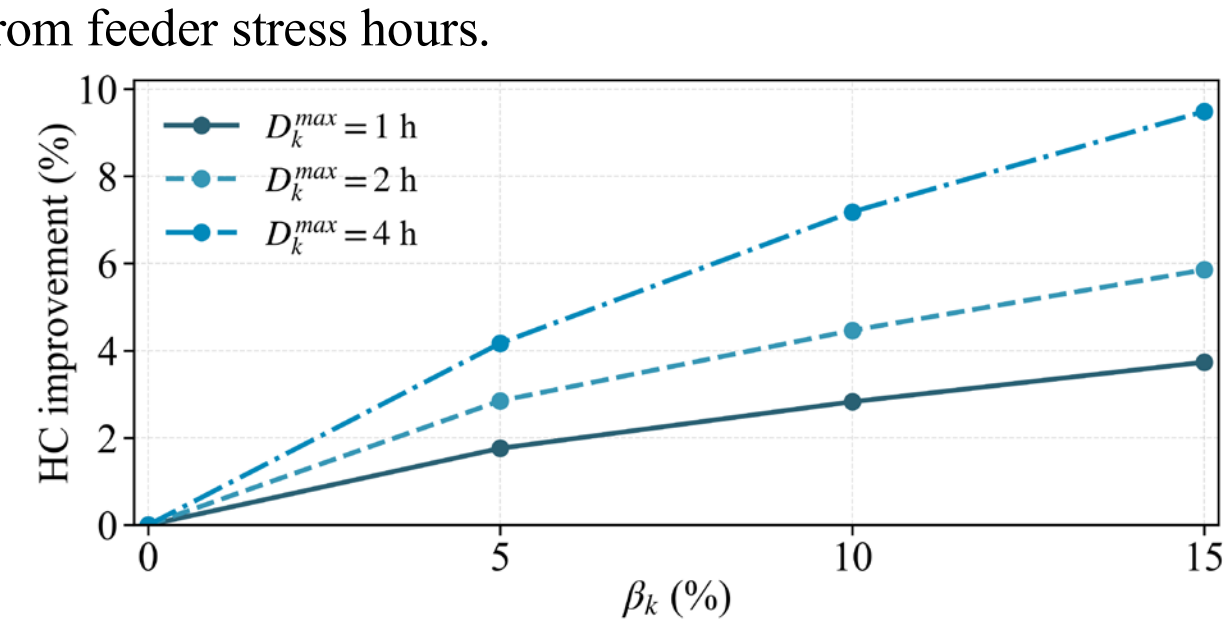

Fig. 2. MEC DC HC improvement under different workload flexibility settings.

TABLE I. REPRESENTATIVE EDGE DC WORKLOAD PROFILES AND FLEXIBILITY ASSUMPTIONS.

| **Archetype** | **Service load profile** | **Shiftable workload arrival profile** | **$\beta_k$ (%) Main / Sensitivity** | **$D_k^{\max}$ (h) Main / Sensitivity** |
|---|---|---|---|---|
| MEC DC | Traffic driven diurnal | Traffic correlated | 10% / 5–15% | 2 h / 1–4 h |
| CDN DC | Content serving | Scheduled cache windows | 15% / 5–20% | 8 h / 4–12 h |
| AI DC | Inference and batch | Hybrid correlated and after hours | 20% / 10–30% | 12 h / 4–12 h |

Fig. 3 illustrates the load reshaping mechanism at Bus 2. Increasing $\beta_k$ shifts more background workload, while increasing $D_k^{\max}$ spreads the execution over a wider time window. However, the optimized facility load remains close to the original profile. This is because the shiftable workload fraction is relatively small, the allowable delay window is short, and the shiftable workload arrival profile is closely correlated with the traffic driven demand peak. Therefore, MEC workload scheduling provides a moderate HC improvement but does not create a large peak reduction.

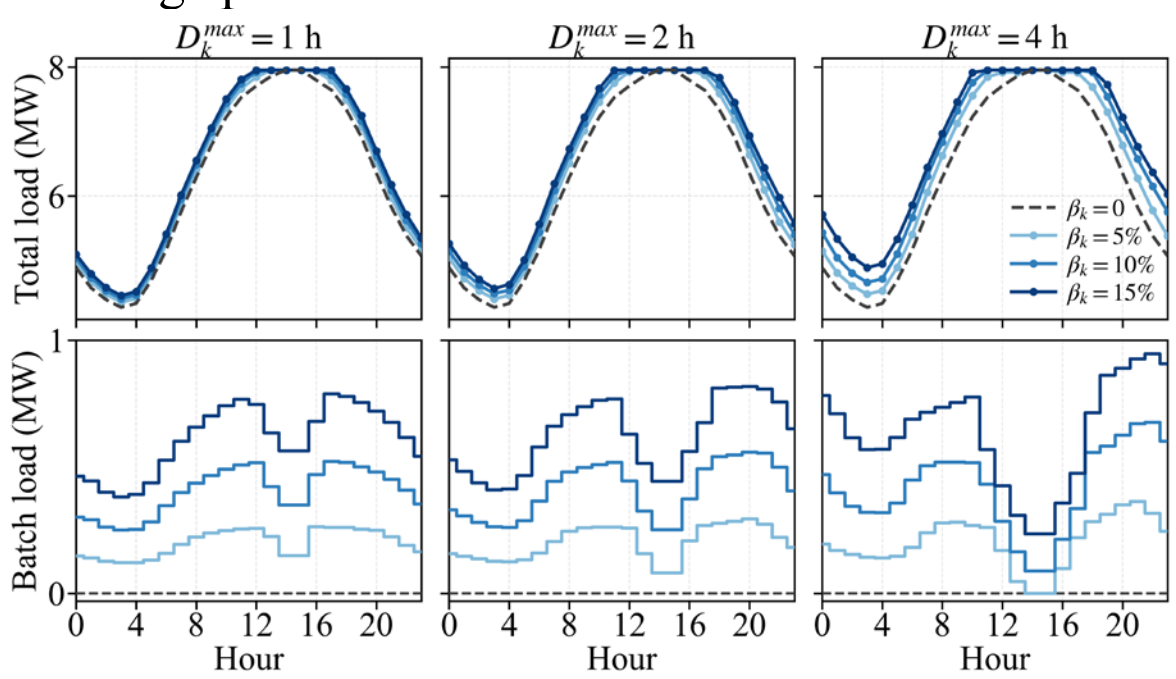

Fig. 3. MEC DC load reshaping and shiftable workload execution at Bus 2.

## C. CDN DC Results: Cache Management Flexibility

The CDN DC case shows a different response. Under the assumed cache management windows, the HC improvement remains negligible across the tested $\beta_k$ and $D_k^{\max}$ settings. This does not mean that the scheduling constraints are inactive. Instead, the schedulable cache management tasks are weakly aligned with the feeder binding period.

Fig. 4 illustrates this mechanism at Bus 2 with $\beta_k$ = 15% and $D_k^{\max}$ = 12 h. The optimized load differs from the original profile, confirming that background cache tasks are scheduled. However, the facility load upper envelope is not materially reduced because the content serving peak remains largely unchanged. Therefore, the apparent limiting peak receives limited relief, and the HC limit remains nearly unchanged.

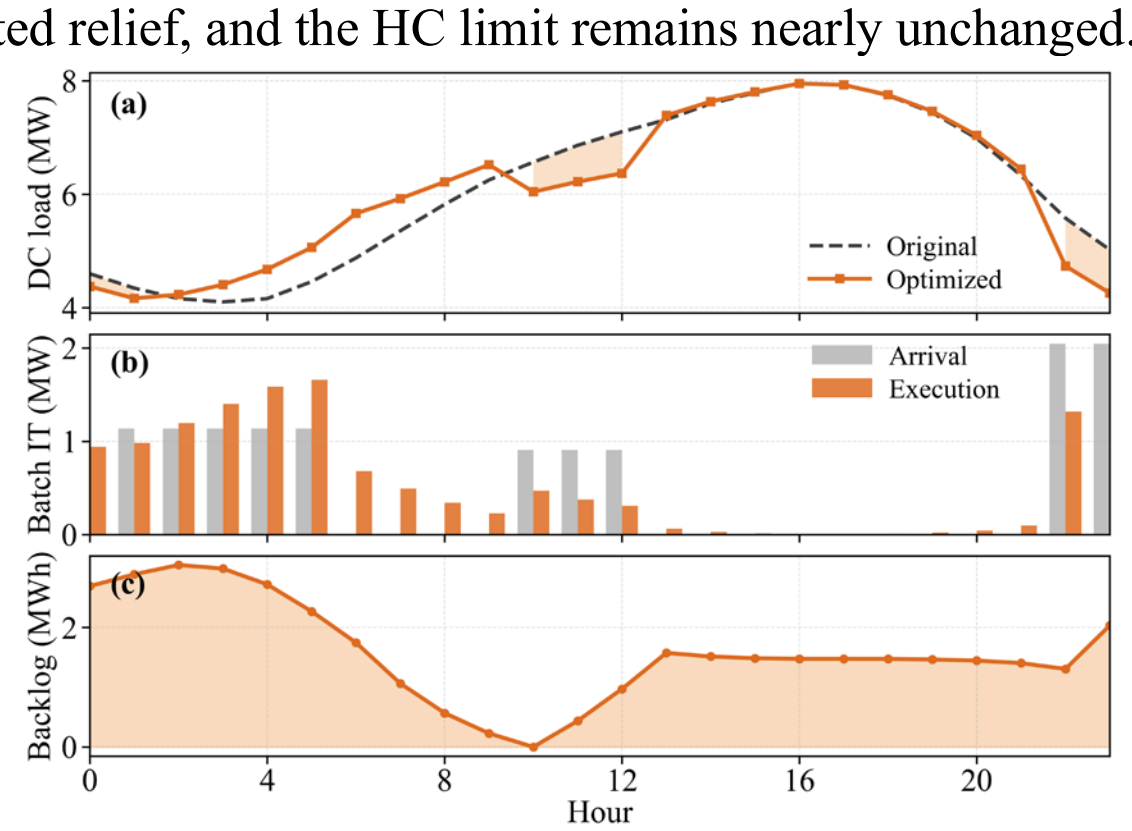

Fig. 4. Illustrative CDN DC workload scheduling at Bus 2 with $\beta_k$ =15%, $D_k^{max}$=12 h: (a) total facility load, (b) batch arrival and execution, and (c) batch backlog.

The shiftable workload arrival, execution, and backlog profiles further confirm that workload conservation and delay constraints are satisfied. The shifted execution mainly occurs during nonbinding or compensation hours rather than during the critical peak period. Therefore, CDN DC flexibility can reshape background workload execution, but it does not necessarily improve HC unless the shiftable workload overlaps with network stress hours. Co-optimizing windows for CDN cache management with periods of feeder stress is left for future work.

## D. AI DC Results and Cross Type Comparison

The AI DC case provides the largest HC improvement among the three tested archetypes because a larger fraction of its workload is batch oriented and delay tolerant. Fig. 5 uses Bus 33 as an illustrative feeder end case to show the delay saturation effect. As shown in Fig. 5(a), HC improvement increases with $\beta_k$. When $\beta_k$ is 10%, increasing $D_k^{\max}$ from 4 h to 12 h has limited impact because the available shiftable workload is small. For larger $\beta_k$, increasing $D_k^{\max}$ from 4 h to 8 h gives a clear improvement, while extending the delay window from 8 h to 12 h gives little additional benefit. This indicates that most useful temporal flexibility is already captured by an 8 h scheduling window in this case.

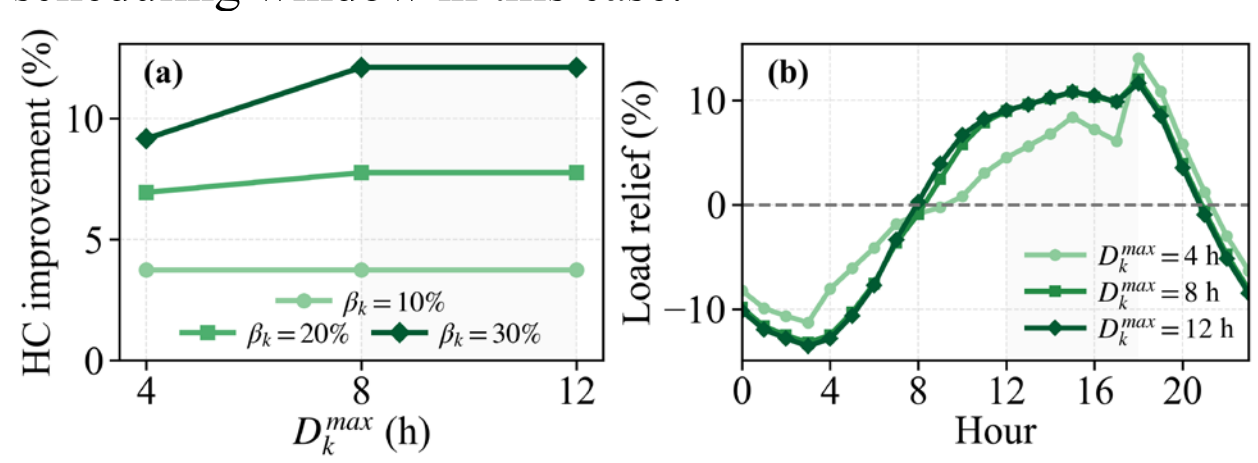

Fig. 5. AI DC HC response and signed load relief at Bus 33: (a) HC gain under different delay limits and (b) signed load relief profiles.

The signed load relief in Fig. 5(b) explains this saturation. Positive values indicate load reduction hours, and negative values indicate compensation hours when deferred workload is executed. Compared with the 4 h case, the 8 h and 12 h cases provide stronger relief during high load hours. However, their relief profiles are similar around the critical period, which explains why the HC gain saturates beyond $D_k^{\max}$ = 8 h.

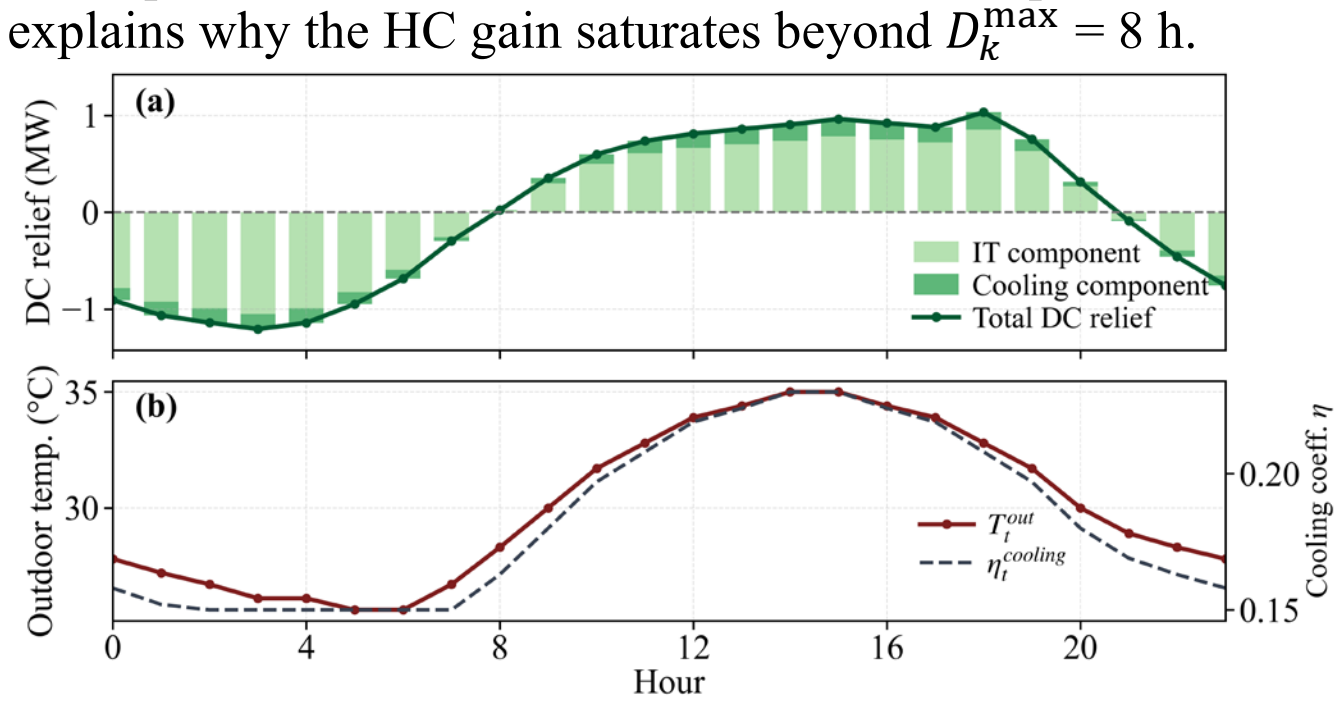

Fig. 6. (a) Cooling aware load relief decomposition for the AI DC case at Bus 2 with $\beta_k$ =15% and $D_k^{max}$=12 h under (b) the adopted daily temperature and cooling coefficient profiles.

Fig. 6 further illustrates the cooling aware load relief mechanism for the AI DC case at Bus 2 when $\beta_k$ is 15% and $D_k^{\max}$ is 12 h. The total DC load relief is decomposed into direct

IT load relief and the associated cooling component. In this case, the outdoor temperature profile and the hourly cooling coefficient are fixed inputs. When workload is shifted away from hours with a higher cooling coefficient, the facility level relief includes both reduced IT demand and reduced cooling related demand. Conversely, executing deferred workload during hours with a lower cooling coefficient leads to a smaller cooling penalty. This result shows that workload flexibility should be evaluated at the facility load level rather than through IT demand alone.

Table II summarizes the main scenario results. Average delay and delayed shiftable workload percentage are computed by FIFO matching of the optimized execution profile to the shiftable workload arrival profile. MEC DCs achieve a maximum HC gain of 4.46% at bus 16. Under the assumed cache management windows, CDN DCs show almost no HC gain because the scheduled cache tasks provide limited relief during the main high load period. AI DCs achieve the largest gain of 9.89% at bus 14, supported by a higher shiftable workload fraction and a longer effective scheduling window. These results show that HC improvement depends not only on the amount of flexible workload, but also on the alignment among workload arrival, delay tolerance, cooling response, and network binding period.

TABLE II. SUMMARY OF HC IMPROVEMENT AND WORKLOAD SCHEDULING PERFORMANCE ON THE IEEE 33-BUS SYSTEM.

| DC type | Main scenario | Max HC gain / Bus | Peak reduction | Average delay | Delayed workload | Peak workload backlog |
|---|---|---|---|---|---|---|
| MEC DC | $\boldsymbol{\beta_k}$=10%, $\boldsymbol{D_k^{max}}$=2 h | 4.46% / Bus 16 | 0.128 MW | 1.02 h | 59.0% | 0.38 MWh |
| CDN DC | $\boldsymbol{\beta_k}$=15%, $\boldsymbol{D_k^{max}}$=8 h | <0.01% / Bus 11 | <0.001 MW | 2.08 h | 61.9% | 1.30 MWh |
| AI DC | $\boldsymbol{\beta_k}$=20%, $\boldsymbol{D_k^{max}}$=12 h | 9.89% / Bus 14 | 0.284 MW | 4.12 h | 75.2% | 2.47 MWh |

## IV. CONCLUSION

This paper presents a HC assessment framework for edge DCs with shiftable workloads. The framework represents three edge DC types, including MEC, CDN, and AI DCs, and converts representative service profiles, prescribed cooling-overhead coefficient, and delay constrained shiftable workload scheduling into grid connected active and reactive loads. The case studies show that workload flexibility provides different HC benefits depending on service type and workload timing. MEC DCs achieve moderate improvement through short delay background workload shifting, but CDN DCs show negligible gain under the assumed cache management windows because their shiftable tasks provide limited relief during the main high load period. Among the modeled archetypes, AI DCs achieve the largest improvement due to the assumed higher shiftable workload fraction and longer effective scheduling window. These results suggest that internal workload scheduling can provide useful load side flexibility for edge DC integration under the considered single feeder operating conditions.

## ACKNOWLEDGMENT

This research was supported in part by the University of Houston Energy Transition Institute Seed Grant Program.